\documentclass[conference]{IEEEtran}

\usepackage{cite}
\usepackage{amsmath,amssymb,amsfonts}
\usepackage{algorithmic}
\usepackage{graphicx}
\usepackage{textcomp}
\usepackage{xcolor}
\usepackage[format=plain,justification=justified]{caption}
\usepackage{mathtools}
\usepackage{subcaption}
\usepackage{hyperref}
\usepackage{hyperref}
\hypersetup{colorlinks,allcolors=black}
\usepackage[section]{placeins}
\usepackage{upgreek} 
\usepackage{amssymb} 
\usepackage{tabularx} 

\def\BibTeX{{\rm B\kern-.05em{\sc i\kern-.025em b}\kern-.08em
    T\kern-.1667em\lower.7ex\hbox{E}\kern-.125emX}}
\begin{document}

\title{Zero noise extrapolation on logical qubits by scaling the error correction code distance
}

\author{\IEEEauthorblockN{Misty A. Wahl}
\IEEEauthorblockA{\textit{Unitary Fund} \\
San Francisco, California 94104, USA \\
misty@unitary.fund}
\and
\IEEEauthorblockN{Andrea Mari}
\IEEEauthorblockA{\textit{Unitary Fund} \\
San Francisco, California 94104, USA \\
}
\IEEEauthorblockA{\textit{Physics Division, School of Science and Technology} \\
\textit{Universit\`a di Camerino}\\
62032 Camerino, Italy\\
andrea@unitary.fund}
\and
\IEEEauthorblockN{Nathan Shammah}
\IEEEauthorblockA{\textit{Unitary Fund} \\
San Francisco, California 94104, USA \\
nathan@unitary.fund}

\and
\IEEEauthorblockN{William J. Zeng}
\IEEEauthorblockA{
\textit{Unitary Fund} \\
San Francisco, California 94104, USA}
\IEEEauthorblockA{
\textit{Quantonation} \\
75010 Paris, France\\
will@unitary.fund}
\and
\IEEEauthorblockN{Gokul Subramanian Ravi}
\IEEEauthorblockA{
\textit{University of Chicago}\\
Chicago, IL 60637, USA\\
}
\IEEEauthorblockA{
\textit{University of Michigan}\\
Ann Arbor, MI 48109, USA\\
gsravi@umich.edu}
}

\maketitle

\begin{abstract}
In this work, we migrate the quantum error mitigation technique of Zero-Noise Extrapolation (ZNE) to fault-tolerant quantum computing. We employ ZNE on \emph{logically encoded} qubits rather than \emph{physical} qubits. This approach will be useful in a regime where quantum error correction (QEC) is implementable but the number of qubits available for QEC is limited. Apart from  illustrating the utility of a traditional ZNE approach (circuit-level unitary folding) for the QEC regime, we propose a novel noise scaling ZNE method specifically tailored to QEC: \emph{distance scaled ZNE (DS-ZNE)}. DS-ZNE scales the distance of the error correction code, and thereby the resulting logical error rate, and utilizes this code distance as the scaling `knob' for ZNE. Logical qubit error rates are scaled until the maximum achievable code distance for a fixed number of physical qubits, and lower error rates (i.e., effectively higher code distances) are achieved via extrapolation techniques migrated from traditional ZNE. Furthermore, to maximize physical qubit utilization over the ZNE experiments, logical executions at code distances lower than the maximum allowed by the physical qubits on the quantum device are executed in parallel across the device, thereby reducing overall circuit execution costs. 

We validate our proposal with numerical simulation for the surface code and confirm that ZNE lowers the logical error rates and increases the effective code distance beyond the physical capability of the quantum device. For instance, at a physical code distance of 11, the DS-ZNE effective code distance is 17, and at a physical code distance of 13, the DS-ZNE effective code distance is 21. When the proposed technique is compared against unitary folding ZNE under the constraint of a fixed number of executions of the quantum device, DS-ZNE outperforms unitary folding by up to 92\% in terms of the post-ZNE logical error rate.
\end{abstract}


\section{Introduction}
\label{intro}
Quantum computers, while holding promise for solving otherwise intractable problems, are afflicted by noise, limiting their usefulness in the near term. The approach of quantum error correction (QEC) presents a means of addressing the effects of noise but requires larger numbers of qubits than those available on current devices. Progress has been demonstrated through improved quality of physical qubits and physical gates, logical qubit lifetimes, fault-tolerant universal gates, logical error rates, and scaling of the surface code~\cite{Sivak_2023, Ni_2023, Postler_2022, PhysRevLett.129.030501, acharya2022suppressing}, but qubit quality and scaling challenges remain as obstacles to implementing QEC for practical applications.

Recently, a focus has been drawn to quantum error mitigation (QEM) techniques~\cite{endo2021hybrid,cai2022quantum}, which can partially counteract the effect of noise in a quantum computation.

A breadth of QEM techniques have been experimentally demonstrated on noisy quantum devices, such as for zero noise extrapolation (ZNE) \cite{kandala2019error,giurgica2020digital, russo2022testing} and similar post-processing methods~\cite{ravi2022boosting}, probabilistic error cancellation (PEC)~\cite{russo2022testing,berg2022probabilistic}, dynamical decoupling~\cite{acharya2022suppressing} and symmetry-based techniques~ \cite{McClean2017,colless2018computation,mcclean2020decoding,huggins2021virtual}. Although QEM techniques on physical qubits generally require no qubit overhead (as in dynamical decoupling, ZNE and PEC) or a smaller qubit overhead than required for QEC (as in symmetry-based QEM techniques~\cite{koczor2021exponential, huggins2021virtual}), sampling overhead remains a practical limitation for some techniques, particularly for PEC.  

One way of viewing the overheads of QEC and QEM in a unified picture is as a trade between QEC's qubit overhead and QEM's sampling overhead. On the one hand, different QEM techniques have been hybridized to realize the benefit of QEM at a lower circuit sampling overhead~\cite{mari2021extending, 10025519, ferracin2022efficiently}. On the other hand, QEM techniques have been inserted in standard QEC implementations, with the purpose of lowering the physical qubit requirement. Recently, PEC has been theoretically applied on logical qubits to reduce the noise of logical gates and effectively increase the code distance~\cite{piveteau2021error, Suzuki_2022_PRXQ}. 

In this work, we extend the use of QEM to reduce the effective logical error rate at fixed qubit overheads to the case of ZNE. In selecting this QEM technique for our hybridized approach, we note that ZNE does not require knowledge of the noise model beyond estimation of the physical error rate, nor does it incur the sampling overhead associated with PEC. 

\begin{figure}[t]
    \includegraphics[width=\linewidth]{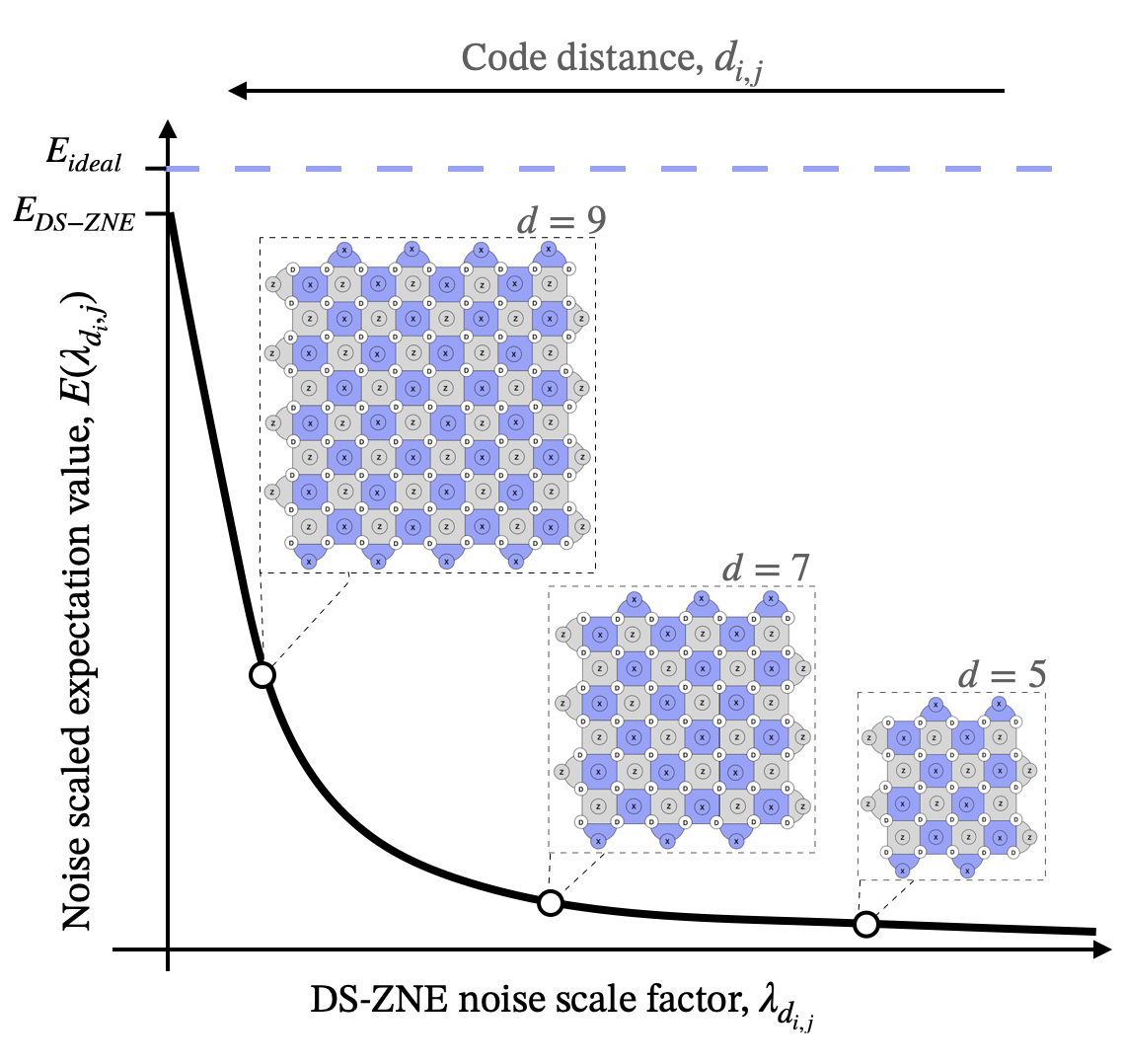}
    \caption{Illustration of noise scaling by code distance as proposed in the DS-ZNE framework. Distance-scaled expectation values $E(\lambda_{d_{i,j}})$ (vertical axis) are evaluated at different noise scale factors $\lambda_{d_{i,j}}$ (horizontal axis) which, in this example are obtained at distances $d = 9$, $d = 7$, and $d = 5$. Note that $\lambda_{d_{i,j}}$ \emph{increases} with \emph{decreasing} $d$. $E_{ideal}$ is the ideal noiseless expectation value (dashed blue horizontal line), and $E_\text{DS-ZNE}$ is the error-mitigated expectation value obtained by a curve fit (solid black curve) and extrapolating to the zero noise limit.}     
    \label{fig:noise-scaling}
\end{figure}

We abstract the level at which ZNE is applied, from the physical circuit level to the logical circuit level. We show that ZNE can mitigate errors in logical qubits through two different noise scaling techniques. First we investigate the effect of noise scaling by global (circuit-level) unitary folding on logical qubits. We then propose and demonstrate a new noise scaling technique for ZNE on logical qubits, which we refer to as \emph{distance scaled ZNE} (DS-ZNE) in which the circuit is executed at higher noise levels by scaling down the code distance. 
While our experiments and evaluation focus on the surface code~\cite{Fowler_2012_PRA}, our results are broadly applicable to other QEC codes as well.
Fig.~\ref{fig:noise-scaling} is a pictorial representation  of the DS-ZNE technique, illustrating  the  relationship between the distance-scaled expectation values $E(\lambda_{d_{i,j}})$, the noise scale factors $\lambda_{d_{i,j}}$, and code distance $d$---a detailed description is provided in Section~\ref{DS-ZNE}.

DZ-ZNE has multiple practical benefits.
First, DS-ZNE is particularly useful for the very realistic scenarios in which the device has a limited number of qubits and the code distance cannot be further increased. Second it is much simpler to implement on devices compared to circuit-folding based ZNE, since changing code distances is somewhat trivial, whereas designing error corrected circuit folded circuits of different depths is not. 
Third, for larger devices, the spared qubits at lower code distances of DS-ZNE can be utilized by allowing additional circuit executions in parallel and thereby improving the results without incurring additional circuit execution costs, not dissimilar to the approach in Ref.~\cite{Combes_2017}. 

Here we propose, and confirm, that expectation values obtained from circuit executions at lower code distances, which are then used to extrapolate to the zero noise limit, can effectively reduce the effect of errors in expectation values obtained on logically encoded qubits. From an error-correction metric perspective, DS-ZNE 
can be seen as increasing the effective code distance as compared to the unmitigated baseline scenario without DS-ZNE.  
For instance, at a physical code distance of 11, the DS-ZNE effective code distance is 17, and at a physical code distance of 13, the DS-ZNE effective code distance is 21. When the proposed technique is compared against unitary folding ZNE under the constraint of a fixed number of executions of the quantum device, DS-ZNE outperforms unitary folding by up to 92\% in terms of the post-ZNE logical error rate.

This work is organized as follows. In Section~\ref{Theory} we give an overview of the key components of QEC and the QEM technique of ZNE to be combined in the DS-ZNE framework. In Section~\ref{DS-ZNE} we describe the DS-ZNE framework and the setup of its demonstration on randomized benchmarking circuits, and in  Section~\ref{Results} we present the results obtained from the demonstration. Finally we conclude and suggest future extensions of the DS-ZNE framework.

\section{Theory}\label{Theory}
\subsection{Distance of Quantum Error Correction Codes}
QEC improves the fidelity of a quantum computation by using additional qubits to detect and correct errors occurring on the physical data qubits~\cite{mike_ike_2020, RevModPhys.87.307, Roffe_2019}. QEC codes encode each logical qubit into an array of physical data qubits. Ancilla qubits are also entangled with each data qubit, allowing the extraction of information about errors without destruction of the quantum state of the data qubits. The output of repeated measurements of the ancilla forms a syndrome, which is a classical error signature. The process of measuring the ancilla maps the errors of the data qubits to discrete Pauli errors, which are often expressed in terms of $X$ (bit-flip) and $Z$ (phase-flip) errors. The syndrome is then passed to a decoder which extracts information from the syndrome about the errors in the data qubits. Correction operations are applied to the data qubits based on the information obtained from the syndrome about the type and location of the errors. 
If the physical qubit error rates are lower than some threshold (which depends on the error code and the decoder), then increasing the size of the physical data array that maps to each logical qubit, will monotonically decrease the logical error rates.
The size of the physical data block per logical qubit is parameterized by the error correction code distance, $d$.  For a code distance of $d$, error chains up to a length of $(d-1)/2$ can be detected and corrected in each error correction cycle, per logical qubit. 

\begin{figure}[ht]
    \includegraphics[width=0.9\linewidth]{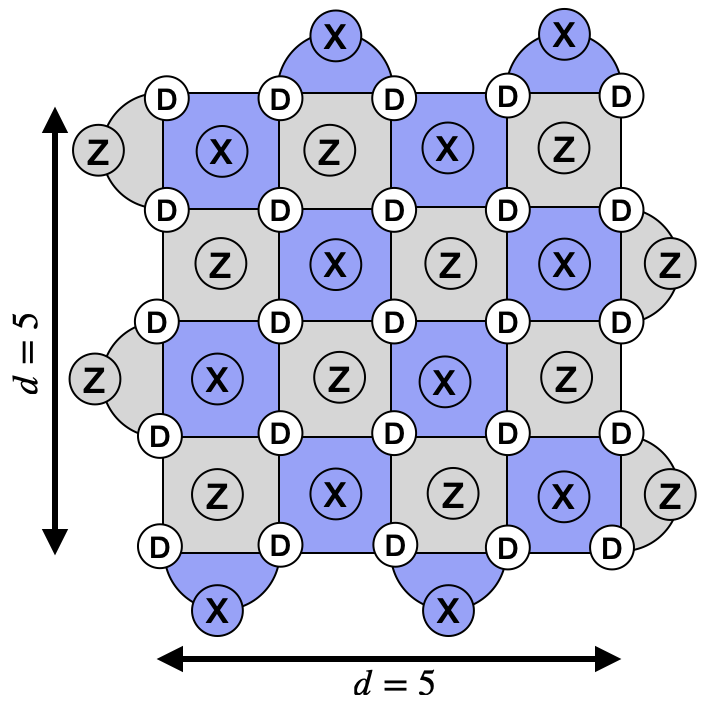}
    \caption{Diagram of the $d = 5$ rotated surface code. The white circles marked "D" represent data qubits and the gray and blue circles represent the ancilla qubits used to measure the stabilizers. Gray squares measure the $Z$ stabilizers and blue squares measure the $X$ stabilizers.}%
    \label{fig:surface-code-background}
\end{figure}

Surface codes are a common choice of error correction code in the immediate future of fault-tolerant quantum computing, as they have high error thresholds (nearly 1\% physical qubit error)~\cite{Fowler_2012_PRA}.
Further, they require only nearest neighbor physical connectivity, employing an alternating pattern of physical data and parity (ancilla) qubits in a 2-dimensional lattice, and are therefore amenable to practical quantum topologies of today.
Fig.~\ref{fig:surface-code-background} shows an illustration of a rotated distance-five surface code. The rotated surface code is condensed version of the surface code, with the benefit of smaller total physical qubit and gate overheads~\cite{bravyi1998quantum, PhysRevA.90.062320, Fowler_2012_PRA, inproceedings}. Data qubits are indicated by white circles and ancilla qubits for stabilizer measurements are indicated by the gray and blue circles. The $Z$ stabilizer measurements are represented by gray squares and the $X$ stabilizer measurements are represented by blue squares. As can be seen in Fig.~\ref{fig:surface-code-background}, the number of physical qubits required to implement an error correction code of distance $d$ is proportional to $d^2$.

We denote the error rate of the physical qubits as $p$ and the threshold as $p_{th}$. As presented in Ref.~\cite{Fowler_2012_PRA}, when operating in the regime of  $p < p_{th}$, the logical error rate $\mathcal{P}_{L}$ decreases exponentially with increasing $d$, and $\mathcal{P}_{L}$ can be approximated by the empirical formula given by
\begin{equation} \label{eq:logical_error_rate}
\mathcal{P}_L\cong 0.03 (p/p_{th})^{(d + 1)/2}.
\end{equation}
Since the logical error rate depends on the code distance, the noise level of the computation can also be scaled by scaling the code distance. This relationship between code distance and noise scaling forms the basis for the DS-ZNE framework.

\subsection{Zero Noise Extrapolation}{\label{ZNE-theory}}
In zero noise extrapolation \cite{temme2017error,li2017efficient,kandala2019error}, a quantum circuit is executed at different noise scale factors $\lambda$ above that of the device to obtain noise-scaled expectation values $E(\lambda)$ of the same observable $A=A^\dag$,
\begin{equation}
E(\lambda)=\text{tr}\left[A\mathcal{U}_\lambda(\rho_0)\right],
\label{e0}
\end{equation}
where $\mathcal{U}_\lambda$ is a quantum channel corresponding to the noisy implementation of the ideal unitary $\mathcal{U}$ at the noise scale factor $\lambda$.

The expectation value $E(\lambda)$ is typically measured at the base hardware noise  ($\lambda=1$) and at multiple higher noise levels ($\lambda >1$). Eventually the ideal noiseless result, i.e.\ $E(0)$, is estimated by fitting a curve to the measured data and extrapolating it to the zero noise limit ($\lambda=0$). A common model for the extrapolating curve is an exponential, given by 
\begin{equation}
E(\lambda) = a_0 + a_1e^{-\lambda a_2}, \qquad a_j \in \mathbb R, \quad a_2\ge 0, \label{eq:exp_extrapolation}
\end{equation}
or a polynomial, given by
\begin{equation}
E(\lambda) = a_0 + a_1\lambda + ... + a_n\lambda^n, \qquad a_j \in \mathbb R.
\label{eq:poly_extrapolation}
\end{equation}
The polynomial model of~\eqref{eq:poly_extrapolation} includes the special cases of linear extrapolation ($n=1$) and Richardson extrapolation ($n + 1=$ number of noise scale factors).

ZNE was formulated and tested on hardware at the pulse level~\cite{temme2017error, kandala2019error} and it has  been abstracted and demonstrated on hardware at the gate level in a form known as digital ZNE~\cite{li2017efficient,he2020resource,giurgica2020digital,larose2022mitiq, russo2022testing}. Due to its simplicity of employment, ZNE has been applied on a variety of studies, including hardware benchmarking tasks, variational algorithms, and quantum simulation~\cite{russo2022testing, larose2022error, schultz2022reducing, 9773204, PhysRevD.106.094502}. A key application requirement for ZNE is the availability of an expectation value, such as the probability of measuring a bitstring of interest as the output of a benchmarking task. 

In digital ZNE~\cite{he2020resource,giurgica2020digital,larose2022error}, noise scaling is abstracted at the circuit or gate level by unitary folding or identity insertion. Unitary folding maps the operations in the circuit of interest such that they are followed by their inverse and then repeated. In the absence of noise the additional operations introduced do not affect the final result, but in the presence of noise they increase the noise level in the calculation. Unitary folding can be performed on individual gates, layers of gates, or on the entire circuit. In the case of entire circuit folding, also known as global folding, the scaling takes the form of
\begin{equation}
\mathcal{U} \rightarrow \mathcal{U}_{\lambda_n} = \mathcal{U}(\mathcal{U} ^ \dag \mathcal{U}) ^{(\lambda_n  -1)/2 }=\mathcal{U}(\mathcal{U} ^ \dag \mathcal{U}) ^ n,
\label{globalfolding}
\end{equation}
where $\lambda_n = 1 + 2n$ and $n=0,1,2,\dots$, corresponding to the noise scale factor and the associated number of $\mathcal U^\dag \mathcal U$ insertions, respectively.  

To provide a fair comparison between mitigated and unmitigated values when analyzing the improvement obtained from QEM, it is helpful to establish a fixed sampling budget, in which the total number of circuit executions is kept constant for the unmitigated and mitigated cases~\cite{giurgica2020digital, russo2022testing}. In the case of ZNE, a simple way of calculating the sampling budget $N_\text{samples}$ is by multiplying the number of circuit instances $N_\text{circ}$ by the number of executions of each circuit instance $N_\text{shots}$, i.e.
\begin{equation}
    N_\text{samples} = N_\text{circ} N_\text{shots}\label{samplingbudget}.
\end{equation}

\section{Distance-scaled ZNE framework} \label{DS-ZNE}
\subsection{Noise scaling by code distance reduction}

In the proposed framework, we combine QEC and QEM by applying ZNE on logical qubits with two different noise scaling methods, unitary folding and code distance scaling. To preserve the structure of the circuit acting on logical qubits, we perform global circuit folding instead of locally folding individual gates. Although implementing unitary folding on error-corrected qubits is not trivial, we assume that it is possible to implement at the circuit level without additional physical qubit overhead. The circuit-level implementation of fault tolerant unitary folding is beyond the scope of this work---here we primarily focus on the simple and very effective distance-scaled DS-ZNE approach.

The DS-ZNE method consists of scaling the noise level of the computation by executing the quantum circuit on the logical qubits at successively lower code distances. Specifically, we parameterize the code distance in terms of two positive integers $i$ and $j$ as follows:
\begin{align}
d_{i, j} = i - j, \qquad  j \in \{j_1,\dots, j_k \}, \label{distance_set} 
\end{align}
where $i=d_{i,0}$ is the maximum distance and $j$ quantifies the distance reduction\footnote{
Note that $d_{i, j}$ must be compatible with the underlying error correcting code. For the surface code considered in this work, we choose $d_{i, j}$ to be odd.
}.
The corresponding noise scale factors are:
\begin{align}
\lambda_{d_{i,j}} = \frac{\left.{\mathcal{P}_L} \right|_{d_{i, j}}}{\left. {\mathcal{P}_L} \right|_{d_{i,0}}}\ge 1, \qquad  j \in \{j_1,\dots, j_k \},\label{noise_scaling_set}
\end{align}
where $\mathcal P_L$ is the logical error rate defined in~\eqref{eq:logical_error_rate}.
It should be noted that $\lambda_{d_{i,j}}$ \emph{increases} with \emph{decreasing} $d_{i,j}$. In this case, to evaluate ~\eqref{e0}, instead of applying the unitary folding formula~\eqref{globalfolding} we use: 
\begin{equation}
\mathcal U \rightarrow \mathcal{U}_{{\lambda}_{d_{i,j}}},
\label{eq:distance-scaling}
\end{equation}
where $\mathcal{U}_{{\lambda}_{d_{i,j}}}$ refers to the noisy implementation of the error corrected unitary with reduced code distance $d_{i,j}$, assuming a maximum available distance of $d_{i,0}$.
In the remaining part of the DS-ZNE framework, as in conventional ZNE, the expectation values obtained at the noise scaled values are fit to a curve and extrapolated to the zero noise limit, yielding an error-mitigated expectation value $E_\text{DS-ZNE}$.
Fig.~\ref{fig:noise-scaling}, a pictorial representation  of the DS-ZNE technique, was presented earlier in Section~\ref{intro}.

\subsection{Parallelization of logical circuits}

\begin{figure}[ht]
    \includegraphics[width=\linewidth]{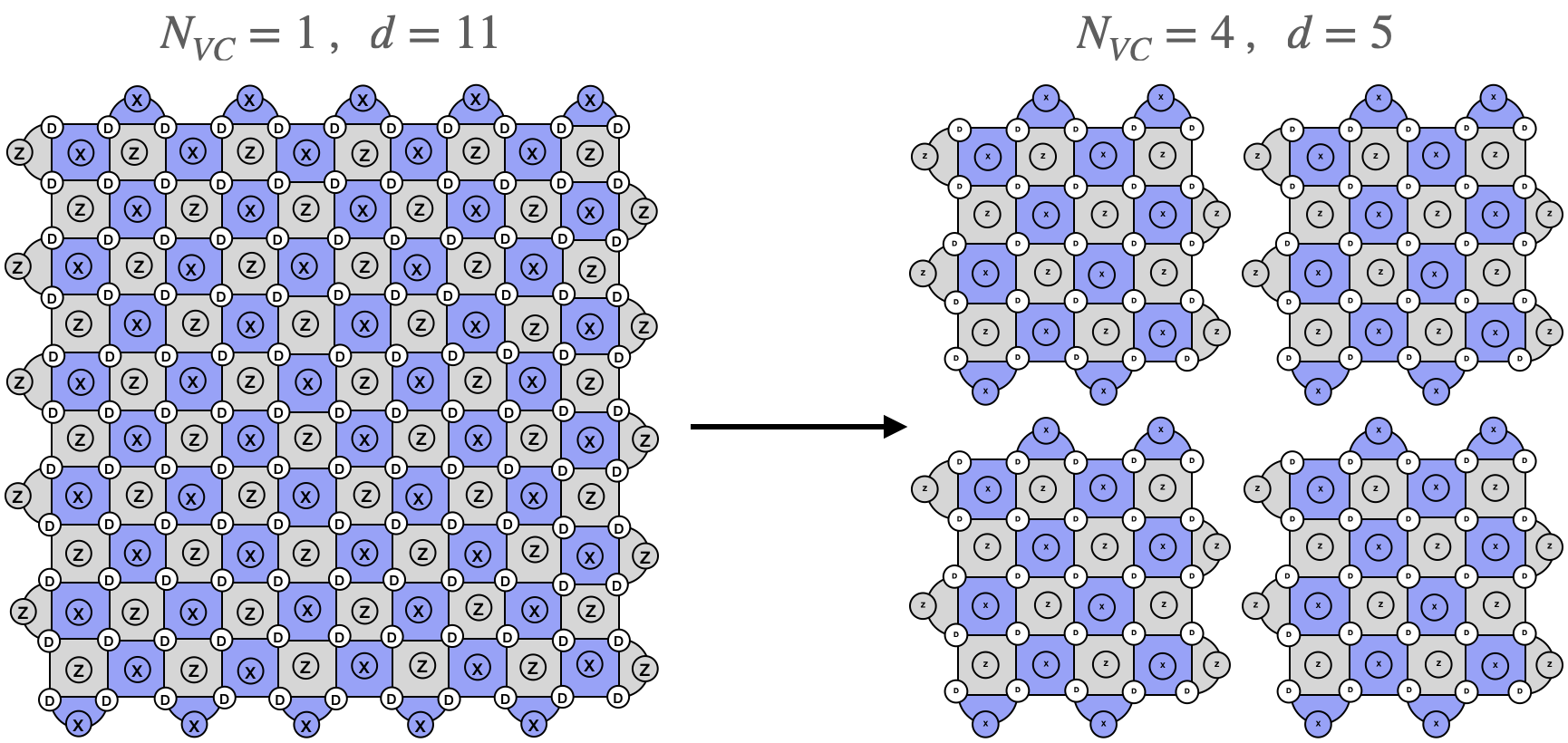}
    \par\bigskip
    \caption{Reducing code distance to increase noise has the added benefit of freeing up qubits that can be reused. In this example, a single qubit in the distance 11 code can be replaced with 4 distance 5 qubits. At scale, this allows the smaller distances to act like virtual processor cores that can be run in parallel for faster sample collection. Thus, distance scaling can use physical qubits more efficiently than other noise scaling techniques, improving performance.
}
    \label{fig:parallelize}
\end{figure}

We now consider how distance scaling enables parallelization of circuit executions within a fixed sampling budget.
At smaller code distances, qubits that are not used in error correction can be re-purposed to behave as multiple virtual processor cores, which are parallel computing regions, similar to those described in Ref.~\cite{PhysRevApplied.18.044064}. We refer to the number of virtual cores as $N_\text{VC}$. The diagram in Fig.~\ref{fig:parallelize} illustrates how such groups of qubits can be run in parallel to collect samples more efficiently, enhancing the performance of the distance scaling technique. For example, given a budget of $N$ circuit executions at code distance $d = 11$, the effective number of measurement shots can be increased from $N$ up to $4 N$, when reducing the code distance from $d=11$ to $d=5$. 
Thus, when mitigating errors with ZNE, distance scaling can use physical qubits more efficiently than other noise scaling techniques, improving the overall performance of the full error mitigation protocol. 

For the particular case of the rotated surface code~\cite{bravyi1998quantum,Dennis_2002,Fowler_2012_PRA}, the number of parallel virtual cores that we can run simultaneously, $N_\text{VC}$, when reducing the distance from $d$ to $d'$, with $d'<d$ is given by

\begin{equation}
N_\text{VC} = \left[\frac{d^2}{d'^2}\right],  
\label{dreduction}
\end{equation}
where $[\cdot]$ represents the integer part. 

This number will vary depending on the error correction code of choice.
It is worth noting that the above estimation only focuses on a single logical qubit. 
The number of possible parallel virtual cores will also depend on how resource overheads such as the physical routing space between logical qubits to perform multi-qubit operations, the physical qubits reserved for magic-state distillation (for T-gates), etc., scale with target error rates. 
Incorporating these resource overheads in the above analysis deserves further exploration but is beyond our current scope.

\begin{figure}[ht]

    \includegraphics[clip, trim=5.4cm 13.5cm 4.2cm 4.5cm, width=\linewidth]{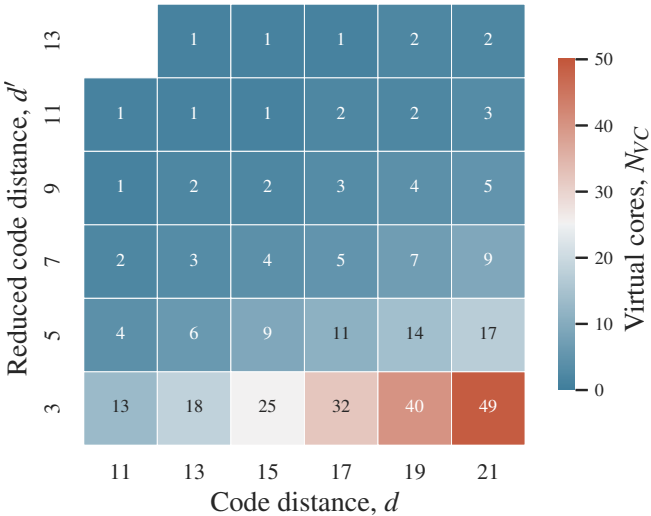}
    \caption{Scaling the distance of a rotated surface code from the original code distance $d$ to a reduced code distance $d'$ corresponds to a gain in virtual cores $N_\text{VC}$, given by~\eqref{dreduction}.
}
    \label{fig:heatmap}
\end{figure}

In the plot of Fig.~\ref{fig:heatmap} we provide an illustration of the gain in virtual cores, $N_\text{VC}$, as a function of $d$ and $d'$, for the specific case of a rotated surface code, as given by ~\eqref{dreduction}. We choose a parameter regime that is believed to remain relevant for near-term quantum devices \cite{Suzuki_2022_PRXQ}.
For a fixed number of qubits, the maximum code distance is fixed for a single virtual core, while $N_\text{VC}$ increases monotonically as $d'$ decreases. However, one should not just optimize for the maximum number of virtual cores, since for a fixed physical error rate, the reduced code distance $d'$ may provide a too noisy expectation value. For this reason, we report our numerical study of DS-ZNE in the next section.

\section{Numerical example based on randomized benchmarking circuits}

\begin{figure}[ht]
    \includegraphics[width=\linewidth]{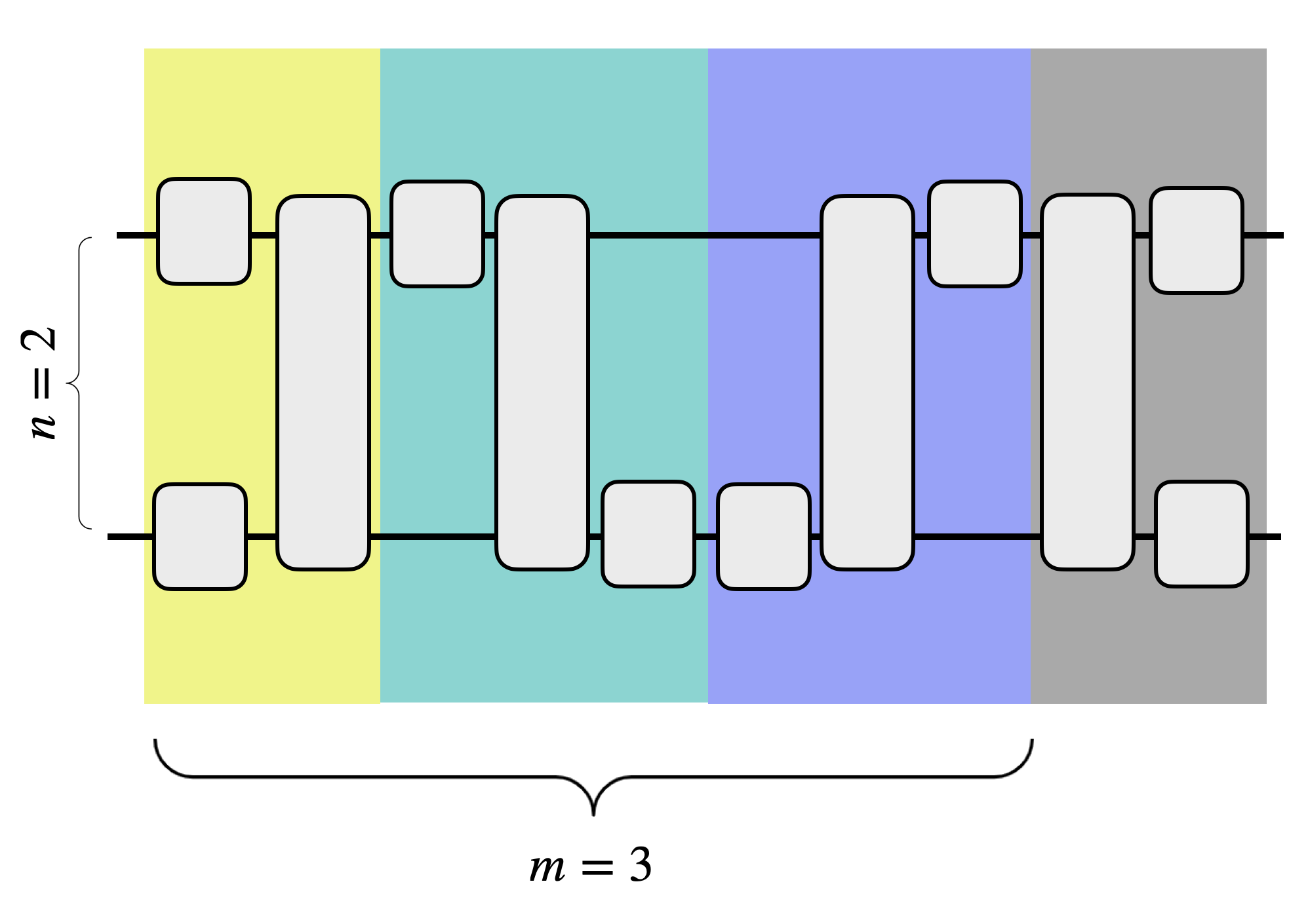}
    \caption{An $n$-qubit, randomized benchmarking circuit of Clifford depth $m$ consists of a random sequence of $m$ elements (represented by the colored rectangles) of the $n$-qubit Clifford group (with operations represented by the white tiles) followed by an inverse (represented by the gray rectangle) to obtain the final state, $|00\rangle$ in the example shown above.}
    \label{fig:randomized_benchmarking}
\end{figure}

We showcase an example of ZNE on error-corrected two-qubit randomized benchmarking circuits. Since the task of randomized benchmarking is characterizing quantum devices and estimating gate errors, it is well-suited to our purpose of demonstrating the DS-ZNE framework. Randomized benchmarking circuits are comprised of random sequences of $m$ elements of the $n$-qubit Clifford group followed by a final inverse element such that, in the absence of noise, the final state is equal to the input state~\cite{emerson2005scalable, PhysRevA.77.012307, PhysRevLett.102.090502}. A randomized benchmarking circuit is pictorially represented in Fig.~\ref{fig:randomized_benchmarking} for the case of $n=2$ and $m=3$, with the colored rectangles representing the Clifford sequences, the white tiles representing circuit operations, and the gray rectangle representing the inverse. In this example the final state is $|00\rangle$. 

Although initially proposed and implemented for characterizing gates applied to physical qubits~\cite{emerson2005scalable, PhysRevA.77.012307, PhysRevLett.102.090502}, randomized benchmarking can also be applied at the logical level~\cite{Combes_2017}. Therefore, it is possible to simulate the action of randomized benchmarking circuits on the logical qubits, instead of simulating the physical qubits directly. In a logical circuit with error correction, the errors remaining after correction are Pauli errors.
We model these errors via single-qubit Pauli operations inserted with probability $\mathcal{P}_{L}$ after every correction cycle, where we assume each cycle corresponds to a layer of gates in the circuit. The simulation of code distance scaling is achieved simply by adjusting $d$ in the formula of the logical error rate given in~\eqref{eq:logical_error_rate}. 

 The first set of numerical simulations consisted of exact density matrix simulations of two-qubit randomized benchmarking circuits, whose depth was parameterized by the Clifford depth (the number of Clifford group elements in the circuit), denoted here as $m$. Simulations were performed on 100 circuits with a Clifford depth $m = 20$ and 100 circuits with $m = 30$. We also demonstrated the technique on a limited set of deeper circuits, in which we used Stim \cite{gidney2021stim} to perform a set of stabilizer simulations on 10 circuits with Clifford depth $m = 100$, 10 circuits with $m = 1000$, and 10 circuits with $m = 10,000$.
 
 We calculate the sampling budget as given by~\eqref{samplingbudget}, with the assumption of 10,000 system executions for each of the 4 noise-scaled circuits used for ZNE. All instances of the randomized benchmarking circuits produce an expectation value of 1 in the absence of noise, as the circuits compile to identity and the observable of interest is the probability of obtaining the input state.

First, distance-scaled expectation values $E(\lambda_{d_{i,j}})$ were evaluated for the set of distances and associated scale factors as defined in~\eqref{distance_set} and~\eqref{noise_scaling_set}, for $ 11 \le i \le 27$ and $j \in \{0, 2, 4, 6\}$, corresponding to $4$ distance scalings. The distances $d_{i,j}$ were selected to be in a range considered achievable in the near-term~\cite{Suzuki_2022_PRXQ} and restricted to odd numbers to represent an efficiently constructed lattice~\cite{Fowler_2012_PRA}. The $d_{i,j}$ in this example were selected with a linear spacing, to prevent excessively wide spacing of noise scale factors (since the noise scale factors have an exponential dependence on $d$) and therefore to produce an accurate curve fit and extrapolation. For simplicity the same code distance was used on both qubits and throughout the circuit. The threshold $p_{th}$ was set at 0.009. Also, assuming operation in the fault-tolerant regime, $p$ was chosen such that $p < p_{th}$, $p = 0.006$ and $p=0.004$ in the first and second parts of this example, respectively. The threshold $p_{th}$ and physical error rate $p$ were used to calculate the logical error rate at each distance ${d_{i,j}}$. For the choice of distance scalings, physical error rate, and randomized benchmarking circuits in this example, the distance-scaled expectation values are well-approximated by a third-order polynomial curve, i.e., by~\eqref{eq:poly_extrapolation} with $n=3$. Therefore, a third-order polynomial extrapolation technique (as described in Sec.~\ref{ZNE-theory}) was applied to each set of distance-scaled expectation values to obtain the corresponding zero noise expectation values. 
 
 Second, the unitary folding based ZNE results were obtained from a third-order polynomial extrapolation on expectation values evaluated at distance $d_{i,0}$ for folding noise scale factors $\lambda_n \in \{1, 3, 5, 7\}$. The unitary folding and extrapolation functions were applied using the ZNE module of the software package Mitiq~\cite{larose2022mitiq}.  Results without mitigation were also obtained at each code distance with the same total circuit execution budget as the results with mitigation, i.e., we used 40,000 unmitigated executions. 
 
\section{Results and Discussion} \label{Results}

\subsection{Error-mitigated expectation values}
\begin{figure}[ht]
\begin{subfigure}{0.489\textwidth}
    \includegraphics[clip, trim=5.4cm 14cm 4.2cm 4.6cm, width=\linewidth]{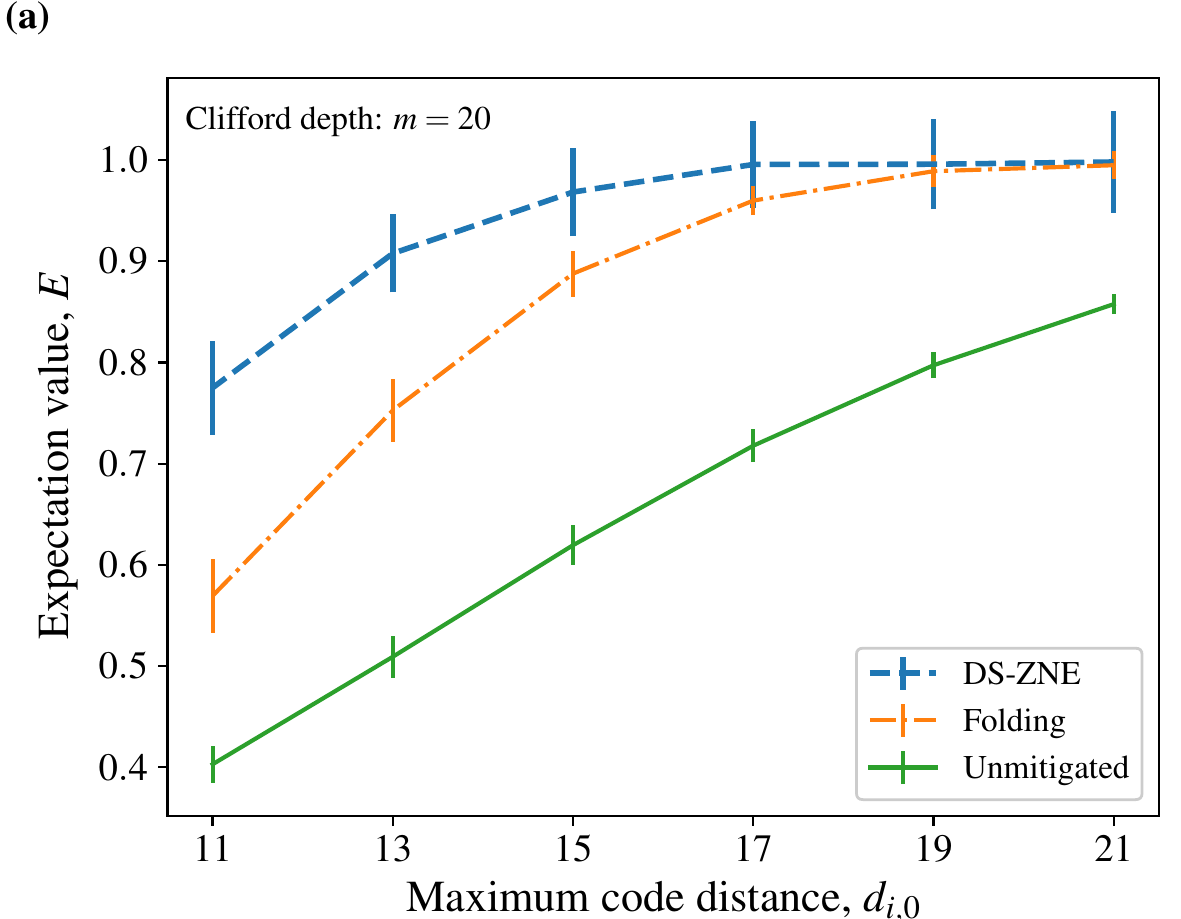}
    \phantomsubcaption
        \label{fig:depth20expvals}
\end{subfigure}
\begin{subfigure}{0.489\textwidth}
    \includegraphics[clip, trim=5.4cm 14cm 4.2cm 4.5cm, width=\linewidth]{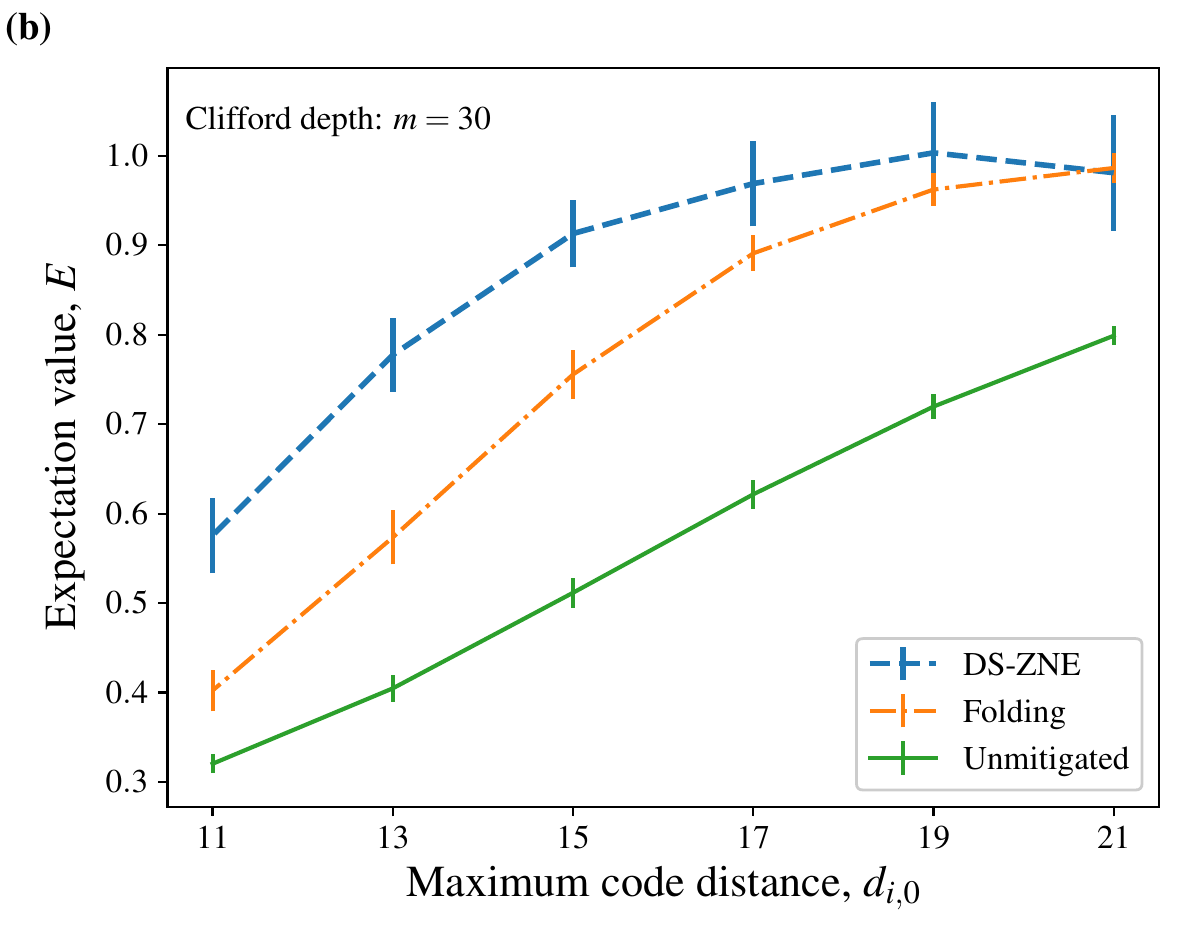}
    \phantomsubcaption
    \label{fig:depth30expvals}
\end{subfigure}
 \caption{Expectation value (vertical axis) of $A=|00\rangle\langle00|$ obtained from randomized benchmarking circuits acting on logical qubits and error-mitigated with DS-ZNE (dashed blue curve), with unitary folding ZNE (dot-dashed orange curve), and also without mitigation (solid green curve).
 The ideal,  noiseless expectation value is 1.
 The horizontal axis $d_{i,0}$ is the highest distance that we assume to be available in a given quantum processor. For DS-ZNE, we used $j\in \{0, 2, 4, 6\}$. For unitary folding ZNE, we used $\lambda_n \in \{1, 3, 5, 7\}$.
 The two plots show the results for
 Clifford depth $ m = 20$ \textbf{(\subref{fig:depth20expvals})} and   $ m = 30$ \textbf{(\subref{fig:depth30expvals})}.}
\label{fig:expvals}
\end{figure}

 The mean expectation value $E$ averaged over 100 trials together with its standard deviation (the error bar) is plotted for each maximum  code distance $d_{i,0}$, at Clifford depths $m =  20$ and $m = 30$, in Fig.~\ref{fig:expvals}. The horizontal axis $d_{i,0}$ is the highest available distance at which the expectation values are evaluated. The dashed blue line represents the error mitigated results based on DS-ZNE with $j \in \{0, 2, 4, 6\}$, the dot-dashed orange line represents the error mitigated results based on unitary folding with $\lambda_n \in \{1, 3, 5, 7\}$, and the solid green line represents the values obtained without error mitigation\footnote{Code and data for the numerical example are available at \url{https://github.com/unitaryfund/research}}. 
 
From the plots of the mean expectation values in Fig.~\ref{fig:expvals} we can see that for every distance $d_{i,0}$, the unmitigated expectation value has a larger bias than that of the error-mitigated expectation values. The bias decreases with increasing code distance for both DS-ZNE and unitary folding as well as for the unmitigated results, which we expect since the logical error rate decreases with increasing code distance. Defining the (mean) effective logical error rate as $ \epsilon = | 1 - \bar{E}|$, we find that unitary folding reduces $\epsilon$ by 96.4\% and 93.1\% at Clifford depths $m = 20$ and $m = 30$ respectively. Moreover, DS-ZNE reduces $\epsilon$ by up to 98.7\% and 98.9\% at $m = 20$ and $m = 30$ respectively. The simulation results indicate that either DS-ZNE or ZNE with unitary folding is effective in mitigating errors in expectation values obtained from  logical circuits.

In terms of the effective logical error rates obtained with DS-ZNE and with unitary folding ZNE, we find that DS-ZNE outperforms unitary folding ZNE by up to 92\%. The benefits from DS-ZNE compared to folding are more pronounced at lower $d_{i,0}$ because there is greater room for improvement, which is representative of challenging applications for QEC with limited numbers of physical qubits. At higher $d_{i,0}$ there is less benefit since a nearly perfect expectation value has already been achieved prior to extrapolation, which is not realistic for critical applications in which the DS-ZNE framework would be employed.

The standard deviation of the error-mitigated expectation values, is up to 6x larger than that of the unmitigated, which may be partly due to all unmitigated samples being taken without noise scaling, i.e. $\lambda = 1$. Also, the standard deviation of the distance-scaled expectation values is up to 5x larger than the standard deviation of expectation values obtained with unitary folding. This may be due to the larger and nonlinear spacing between noise scale factors for distance scaling as compared to unitary folding, which arises from the exponential scaling of the logical error rate with code distance.

\subsection{Effective code distance}\label{effective_code_distance}
The effective logical error rates can be further analyzed in terms of the effective code distance, as in~\cite{Suzuki_2022_PRXQ}. For an error-mitigated code with distance $d$ the effective code distance is the code distance for which an (approximately) equivalent expectation value can be obtained without mitigation. The effective code distance obtained with distance scaling is denoted as $d_{\text{DS}}$, and the effective code distance obtained with folding is denoted as $d_{\text{F}}$. For example, the effective logical error rate obtained at $m = 30$ with distance scaling at code distance 11 is approximately equal to the unmitigated effective logical error rate at code distance 17, resulting in a $d_{\text{DS}}$ of 17. Similarly, the effective logical error rate obtained at $m = 30$ with distance scaling at code distance 13 is approximately equal to the unmitigated effective logical error rate at code distance 21, resulting in a $d_{\text{DS}}$ of 21. From these results, we see that DS-ZNE increases the effective code distance, and the effect strengthens as code distance increases. This means, as shown in Table~\ref{table_qubit_budget}, that ZNE reduces the total number of physical qubits required to reach an equivalent effective logical error rate.  With unitary folding the reduction $\Delta n_{\text{F}}$ is up to 240 physical qubits and with distance scaling the reduction $\Delta n_{\text{DS}}$ is up to 272 qubits. We expect that the benefits of DS-ZNE will only improve further for more complex applications (as they will need greater sampling budgets) and for greater maximum code distances, which, in addition to a lower starting logical error rate, allow for finer tuning of the noise scale factors.

\begin{table}
\begin{tabularx}{\linewidth} { 
  | | >{\centering\arraybackslash}X 
  | >{\centering\arraybackslash}X 
  | >{\centering\arraybackslash}X 
  | >{\centering\arraybackslash}X 
  | >{\centering\arraybackslash}X 
  | >{\centering\arraybackslash}X || }
 \hline
$\textbf{m}$ &$\textbf{d}$ & $d_{\text{F}}$ & $d_{\text{DS}}$ & $\Delta n_{\text{F}}$ & $\Delta n_{\text{DS}}$\\ [0.5ex] 
 \hline\hline
 20 & 11 & 15 & 19 & 104 & 240 \\
 20 & 13 & 19 & 21 & 240 & 272 \\
 30 & 11 & 13 & 17 & 48 & 168 \\
 30 & 13 & 17 & 21 & 120 & 272 \\ [1ex] 
 \hline
\end{tabularx}
\caption{At a Clifford depth $m$, ZNE increases the effective code distance from $d = d_{i,0}$, to $d_{\text{DS}}$ in the case of distance scaling or $d_{\text{F}}$ in the case of unitary folding. The effective code distance $d_{\text{F}}$, corresponds to a reduction in the required number of physical qubits by $\Delta n_{\text{F}}$ and the effective code distance $d_{\text{DS}}$, corresponds to a reduction in the required number of physical qubits by $\Delta n_{\text{DS}}$.}
\label{table_qubit_budget}
\end{table}

We can see that with reuse of unused qubits, distance scaling can increase the effective code distance within a fixed (serial) execution budget. In addition, distance scaling has the advantage that it does not incur additional overhead from increasing the circuit depth, as in unitary folding, and the runtime benefit of DS-ZNE over unitary folding becomes more significant at larger circuit depths. For example, unitary folding with scale factors $\lambda_n \in  \{1, 3, 5, 7\} $ will result in circuits with 1, 3, 5, and 7 times the original circuit depth.

Finer adjustment of the noise scale factors in distance scaling could yield further improvements, both in increasing the effective code distance and reducing the standard deviation in the distance-scaled results. Instead of using a uniform code distance on both qubits and on each layer of the circuit, the technique could be extended by varying the code distance on different qubits, or in different circuit layers or groups of layers. We anticipate that using different code distances at different places in the circuit would result in an overall noise level that is in between those of the uniformly applied higher code distance and the uniformly applied lower code distance. In that case, the spacing of the intervening noise scale factors obtained from non-uniform code distance scaling could be tuned to improve the accuracy of the extrapolation on every set of noise-scaled expectation values. 

\subsection{DS-ZNE on Longer Circuits}
\begin{figure}[ht]
    \begin{subfigure}{0.489\textwidth}        
        \includegraphics[clip, trim=5.4cm 14cm 4.2cm 4.5cm, width=\linewidth]{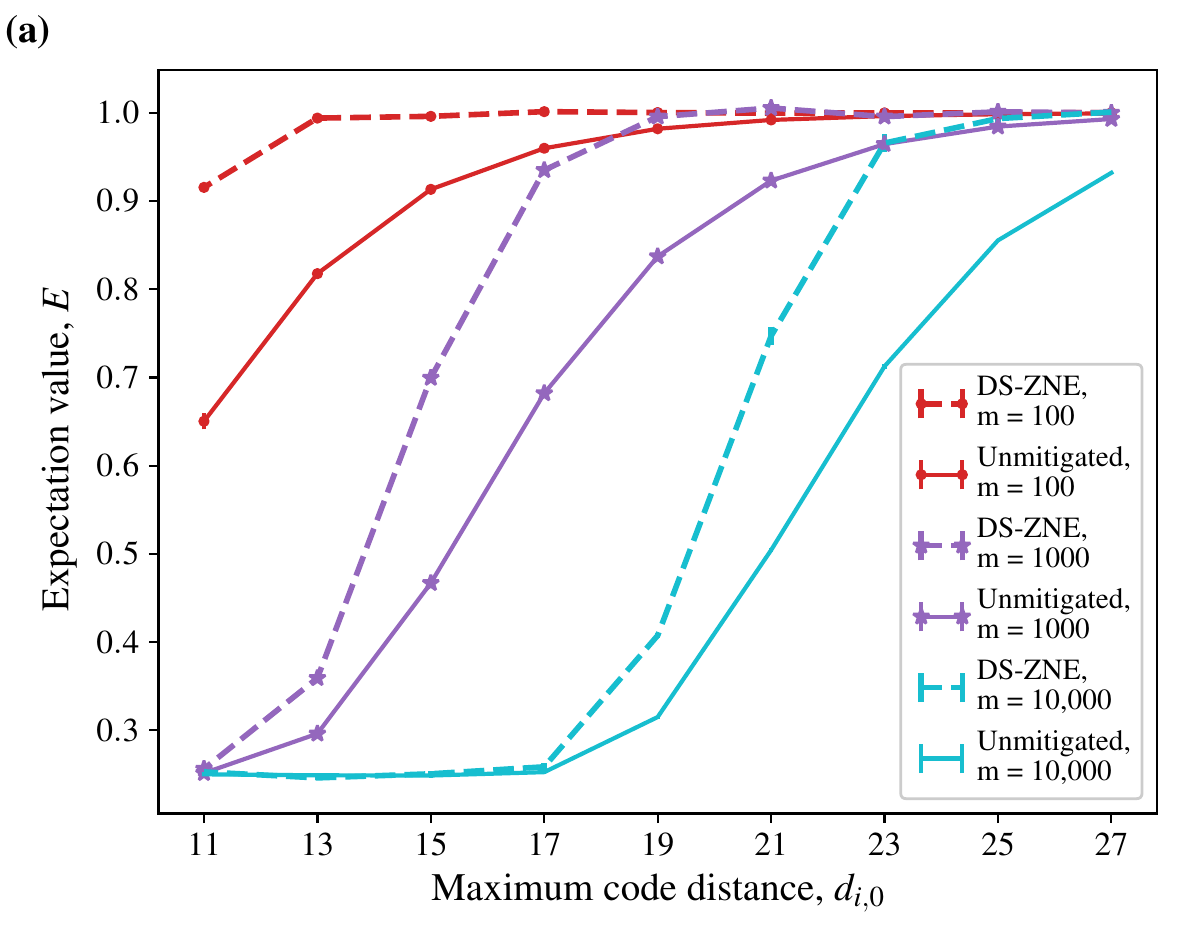}
        \phantomsubcaption
        \label{fig:long_circ_ds_zne_ds_unmit}
    \end{subfigure}
    \begin{subfigure}{0.489\textwidth}
            \includegraphics[clip, trim=5.4cm 14cm 4.2cm 4.5cm, width=\linewidth]{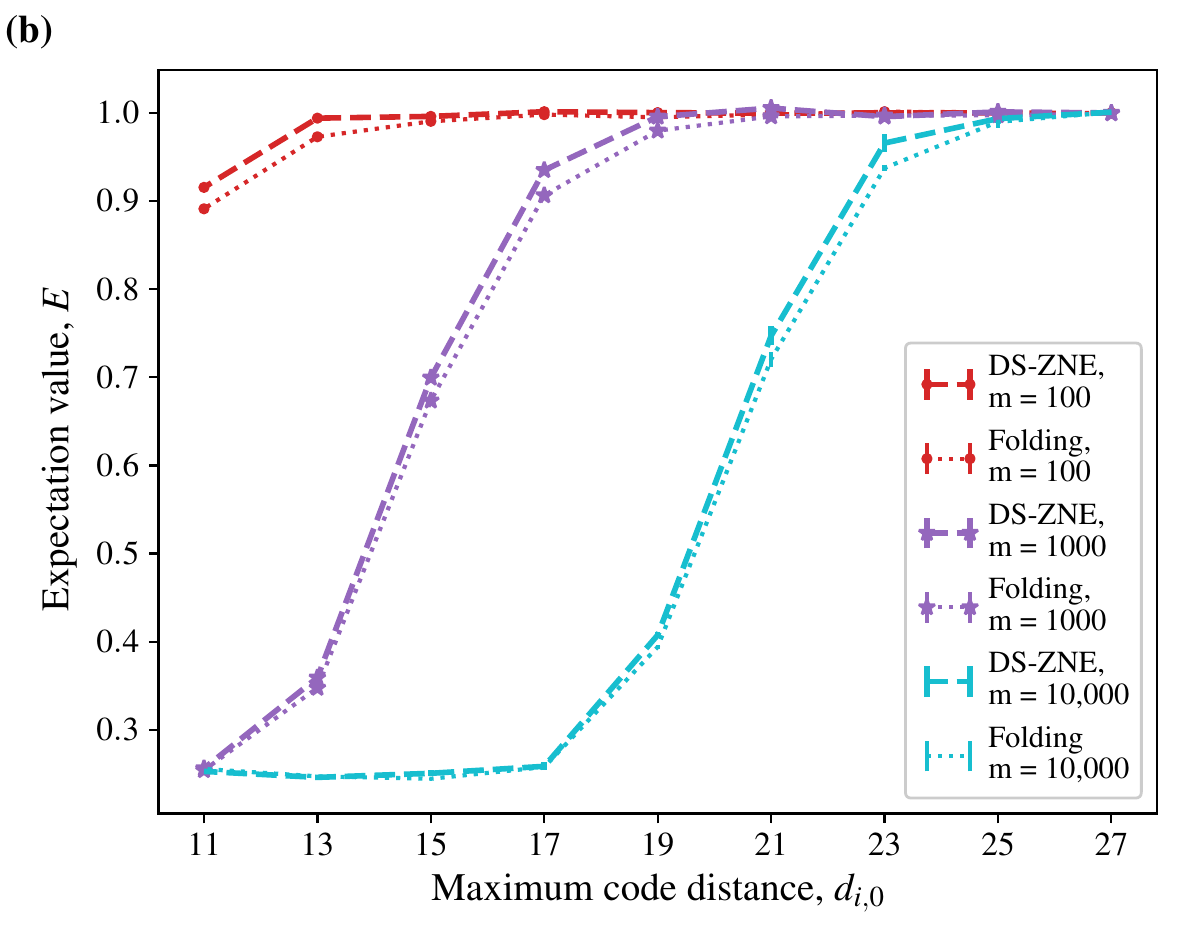}
        \phantomsubcaption
       \label{fig:long_circ_ds_zne_ds_fold}
     \end{subfigure}
    \caption{Expectation value (vertical axis) of $A=|00\rangle\langle00|$ obtained from randomized benchmarking circuits acting on logical qubits and error-mitigated with DS-ZNE of Clifford depths $m = 100$ (red curves with point markers), $m = 1000$ (purple curves with star markers), and $m = 10,000$ (cyan curves), acting on logical qubits and error-mitigated with DS-ZNE (dashed curves),  without mitigation (solid curves in \textbf{(\subref{fig:long_circ_ds_zne_ds_unmit})}), and mitigated with unitary folding (dotted curves in \textbf{(\subref{fig:long_circ_ds_zne_ds_fold})}). The ideal, noiseless expectation value is 1. The horizontal axis $d_{i,0}$ is the highest distance that we assume to be available in a given quantum processor.}
\label{fig:long_circ_ds_zne}
\end{figure}

 The DS-ZNE and unitary folding methods were also demonstrated on a limited set of deeper circuits, indicating the extensibility of the technique to circuits of depths more commensurate with the fault-tolerant regime. In Fig.~\ref{fig:long_circ_ds_zne}, the mean expectation value $E$ averaged over 10 trials, at Clifford depths $m = 100$, $m = 1000$ and $m = 10,000$, is plotted together with its standard deviation (the error bar) for each maximum code distance $d_{i,0}$. The horizontal axis $d_{i,0}$ is the highest available distance at which the expectation values are evaluated. The dashed lines represent the error mitigated results based on DS-ZNE with $j \in \{0, 2, 4, 6\}$, the solid lines represent the values obtained without error mitigation, and the dotted lines represent the error mitigated results based on unitary folding with $\lambda_n \in \{1, 3, 5, 7\}$.

We can see from the plots of the mean expectation values of the higher Clifford depth circuits in Fig.~\ref{fig:long_circ_ds_zne} that for every distance $d_{i,0}$ evaluated at a constant $m$, the unmitigated expectation value has a larger bias than that of the error-mitigated expectation values. Here also the bias decreases with increasing code distance for both DS-ZNE and unitary folding as well as for the unmitigated results, as expected. We find that unitary folding reduces $\epsilon$ by 99.39\%,  99.74\%, and 99.87\%, at Clifford depths $m = 100$, $m = 1000$ and $m = 10,000$ respectively, and DS-ZNE reduces $\epsilon$ by up to 99.98\%,  99.96\%, and 99.85\% at $m = 100$, $m = 1000$ and $m = 10,000$ respectively. The results indicate that even at larger Clifford depths, both techniques are effective in mitigating errors in expectation values obtained from logical circuits.

\section{Conclusion}
We have demonstrated the use of ZNE applied to logical qubits with two different noise scaling methods: with unitary folding and with novel distance scaling. Furthermore, we have shown that ZNE with distance scaling or with circuit-level unitary folding can effectively mitigate errors in expectation values obtained with logically encoded qubits. Equivalently, we can see that ZNE increases the effective code distance for a fixed number of serial circuit executions.  At lower distances, distance scaling outperforms unitary folding, both in terms of the effective logical error rate and in terms of the effective code distance. We anticipate the improvements obtained with ZNE to be even more pronounced for critical applications requiring greater sampling budgets and greater maximum code distances. 

Another benefit of distance scaling over unitary folding is that it does not incur additional overhead in the number of gates in the circuit. Since distance scaling can be applied independently from the type of circuit acting on the encoded qubit, it opens the possibility to use ZNE in applications where unitary folding cannot be applied and a limited amount of qubits are available for error correction. The improvements obtained with ZNE on logical qubits indicate that this combination of QEC and QEM techniques presents a promising method to reduce the effect of errors on the results of the computation, while avoiding prohibitive resource overheads.\\

\section*{Acknowledgment}

This work was supported by the U.S. Department of Energy, Office of Science, Office of Advanced Scientific Computing Research, Accelerated Research in Quantum Computing under Award Number DE-SC0020266 as well as by IBM under Sponsored Research Agreement No. W1975810.
AM acknowledges support from the PNRR
MUR project PE0000023-NQSTI. 
This work is funded in part by EPiQC, an NSF Expedition in Computing, under award CCF-1730449; 
in part by STAQ under award NSF Phy-1818914; in part by NSF award 2110860; 
in part by the US Department of Energy Office  of Advanced Scientific Computing Research, Accelerated Research for Quantum Computing Program; 
and in part by the NSF Quantum Leap Challenge Institute for Hybrid Quantum Architectures and Networks (NSF Award 2016136) 
and in part based upon work supported by the U.S. Department of Energy, Office of Science, National Quantum Information Science Research Centers.  
GSR thanks Frederic Chong and Jonathan Baker for helpful discussions.

\bibliographystyle{IEEEtranS}
\bibliography{refs}

\begin{thebibliography}{10}
\providecommand{\url}[1]{#1}
\csname url@samestyle\endcsname
\providecommand{\newblock}{\relax}
\providecommand{\bibinfo}[2]{#2}
\providecommand{\BIBentrySTDinterwordspacing}{\spaceskip=0pt\relax}
\providecommand{\BIBentryALTinterwordstretchfactor}{4}
\providecommand{\BIBentryALTinterwordspacing}{\spaceskip=\fontdimen2\font plus
\BIBentryALTinterwordstretchfactor\fontdimen3\font minus
  \fontdimen4\font\relax}
\providecommand{\BIBforeignlanguage}[2]{{%
\expandafter\ifx\csname l@#1\endcsname\relax
\typeout{** WARNING: IEEEtranS.bst: No hyphenation pattern has been}%
\typeout{** loaded for the language `#1'. Using the pattern for}%
\typeout{** the default language instead.}%
\else
\language=\csname l@#1\endcsname
\fi
#2}}
\providecommand{\BIBdecl}{\relax}
\BIBdecl

\bibitem{berg2022probabilistic}
\BIBentryALTinterwordspacing
E.~v.~d. Berg, Z.~K. Minev, A.~Kandala, and K.~Temme, ``Probabilistic error
  cancellation with sparse {P}auli-{L}indblad models on noisy quantum
  processors,'' \emph{arXiv preprint arXiv:2201.09866}, 2022. [Online].
  Available:
  \url{https://ui.adsabs.harvard.edu/link_gateway/2022arXiv220109866V/doi:10.48550/arXiv.2201.09866}
\BIBentrySTDinterwordspacing

\bibitem{bravyi1998quantum}
\BIBentryALTinterwordspacing
S.~B. Bravyi and A.~Y. Kitaev, ``Quantum codes on a lattice with boundary,''
  1998. [Online]. Available: \url{https://arxiv.org/abs/quant-ph/9811052}
\BIBentrySTDinterwordspacing

\bibitem{cai2022quantum}
\BIBentryALTinterwordspacing
Z.~Cai, R.~Babbush, S.~C. Benjamin, S.~Endo, W.~J. Huggins, Y.~Li, J.~R.
  McClean, and T.~E. O’Brien, ``Quantum error mitigation,'' \emph{arXiv
  preprint arXiv:2210.00921}, 2022. [Online]. Available:
  \url{https://arxiv.org/abs/2210.00921}
\BIBentrySTDinterwordspacing

\bibitem{PhysRevLett.102.090502}
\BIBentryALTinterwordspacing
J.~M. Chow, J.~M. Gambetta, L.~Tornberg, J.~Koch, L.~S. Bishop, A.~A. Houck,
  B.~R. Johnson, L.~Frunzio, S.~M. Girvin, and R.~J. Schoelkopf, ``Randomized
  benchmarking and process tomography for gate errors in a solid-state qubit,''
  \emph{Phys. Rev. Lett.}, vol. 102, p. 090502, Mar 2009. [Online]. Available:
  \url{https://link.aps.org/doi/10.1103/PhysRevLett.102.090502}
\BIBentrySTDinterwordspacing

\bibitem{colless2018computation}
\BIBentryALTinterwordspacing
J.~I. Colless, V.~V. Ramasesh, D.~Dahlen, M.~S. Blok, M.~E. Kimchi-Schwartz,
  J.~R. McClean, J.~Carter, W.~A. de~Jong, and I.~Siddiqi, ``Computation of
  molecular spectra on a quantum processor with an error-resilient algorithm,''
  \emph{Phys. Rev. X}, vol.~8, p. 011021, Feb 2018. [Online]. Available:
  \url{https://link.aps.org/doi/10.1103/PhysRevX.8.011021}
\BIBentrySTDinterwordspacing

\bibitem{Combes_2017}
\BIBentryALTinterwordspacing
J.~Combes, C.~Granade, C.~Ferrie, and S.~T. Flammia, ``Logical randomized
  benchmarking,'' 2017. [Online]. Available:
  \url{https://doi.org/10.48550/arXiv.1702.03688}
\BIBentrySTDinterwordspacing

\bibitem{Dennis_2002}
\BIBentryALTinterwordspacing
E.~Dennis, A.~Kitaev, A.~Landahl, and J.~Preskill, ``Topological quantum
  memory,'' \emph{Journal of Mathematical Physics}, vol.~43, no.~9, pp.
  4452--4505, Sep 2002. [Online]. Available:
  \url{https://doi.org/10.1063/1.1499754}
\BIBentrySTDinterwordspacing

\bibitem{emerson2005scalable}
J.~Emerson, R.~Alicki, and K.~{\.Z}yczkowski, ``Scalable noise estimation with
  random unitary operators,'' \emph{Journal of Optics B: Quantum and
  Semiclassical Optics}, vol.~7, no.~10, p. S347, 2005.

\bibitem{endo2021hybrid}
\BIBentryALTinterwordspacing
S.~Endo, Z.~Cai, S.~C. Benjamin, and X.~Yuan, ``Hybrid quantum-classical
  algorithms and quantum error mitigation,'' \emph{Journal of the Physical
  Society of Japan}, vol.~90, no.~3, p. 032001, 2021. [Online]. Available:
  \url{https://journals.jps.jp/doi/10.7566/JPSJ.90.032001}
\BIBentrySTDinterwordspacing

\bibitem{ferracin2022efficiently}
\BIBentryALTinterwordspacing
S.~Ferracin, A.~Hashim, J.-L. Ville, R.~Naik, A.~Carignan-Dugas, H.~Qassim,
  A.~Morvan, D.~I. Santiago, I.~Siddiqi, and J.~J. Wallman, ``Efficiently
  improving the performance of noisy quantum computers,'' \emph{arXiv preprint
  arXiv:2201.10672}, 2022. [Online]. Available:
  \url{https://arxiv.org/abs/2201.10672}
\BIBentrySTDinterwordspacing

\bibitem{Fowler_2012_PRA}
\BIBentryALTinterwordspacing
A.~G. Fowler, M.~Mariantoni, J.~M. Martinis, and A.~N. Cleland, ``Surface
  codes: Towards practical large-scale quantum computation,'' \emph{Phys. Rev.
  A}, vol.~86, p. 032324, Sep 2012. [Online]. Available:
  \url{https://link.aps.org/doi/10.1103/PhysRevA.86.032324}
\BIBentrySTDinterwordspacing

\bibitem{gidney2021stim}
\BIBentryALTinterwordspacing
C.~Gidney, ``Stim: a fast stabilizer circuit simulator,'' \emph{{Quantum}},
  vol.~5, p. 497, Jul. 2021. [Online]. Available:
  \url{https://doi.org/10.22331/q-2021-07-06-497}
\BIBentrySTDinterwordspacing

\bibitem{giurgica2020digital}
\BIBentryALTinterwordspacing
T.~Giurgica-Tiron, Y.~Hindy, R.~LaRose, A.~Mari, and W.~J. Zeng, ``Digital zero
  noise extrapolation for quantum error mitigation,'' in \emph{2020 IEEE
  International Conference on Quantum Computing and Engineering (QCE)}.\hskip
  1em plus 0.5em minus 0.4em\relax IEEE, 2020, pp. 306--316. [Online].
  Available: \url{https://ieeexplore.ieee.org/document/9259940}
\BIBentrySTDinterwordspacing

\bibitem{acharya2022suppressing}
\BIBentryALTinterwordspacing
{Google Quantum AI}, ``Suppressing quantum errors by scaling a surface code
  logical qubit,'' \emph{arXiv preprint arXiv:2207.06431}, 2022. [Online].
  Available: \url{https://doi.org/10.1038/s41586-022-05434-1}
\BIBentrySTDinterwordspacing

\bibitem{he2020resource}
\BIBentryALTinterwordspacing
A.~He, B.~Nachman, W.~A. de~Jong, and C.~W. Bauer, ``Zero-noise extrapolation
  for quantum-gate error mitigation with identity insertions,'' \emph{Phys.
  Rev. A}, vol. 102, p. 012426, Jul 2020. [Online]. Available:
  \url{https://link.aps.org/doi/10.1103/PhysRevA.102.012426}
\BIBentrySTDinterwordspacing

\bibitem{PhysRevD.106.094502}
\BIBentryALTinterwordspacing
E.~Huffman, M.~Garc\'{\i}a~Vera, and D.~Banerjee, ``Toward the real-time
  evolution of gauge-invariant ${\mathbb{z}}_{2}$ and $u(1)$ quantum link
  models on noisy intermediate-scale quantum hardware with error mitigation,''
  \emph{Phys. Rev. D}, vol. 106, p. 094502, Nov 2022. [Online]. Available:
  \url{https://link.aps.org/doi/10.1103/PhysRevD.106.094502}
\BIBentrySTDinterwordspacing

\bibitem{huggins2021virtual}
\BIBentryALTinterwordspacing
W.~J. Huggins, S.~McArdle, T.~E. O'Brien, J.~Lee, N.~C. Rubin, S.~Boixo, K.~B.
  Whaley, R.~Babbush, and J.~R. McClean, ``Virtual distillation for quantum
  error mitigation,'' \emph{Phys. Rev. X}, vol.~11, p. 041036, Nov 2021.
  [Online]. Available:
  \url{https://link.aps.org/doi/10.1103/PhysRevX.11.041036}
\BIBentrySTDinterwordspacing

\bibitem{PhysRevApplied.18.044064}
\BIBentryALTinterwordspacing
H.~Jnane, B.~Undseth, Z.~Cai, S.~C. Benjamin, and B.~Koczor, ``Multicore
  quantum computing,'' \emph{Phys. Rev. Appl.}, vol.~18, p. 044064, Oct 2022.
  [Online]. Available:
  \url{https://link.aps.org/doi/10.1103/PhysRevApplied.18.044064}
\BIBentrySTDinterwordspacing

\bibitem{kandala2019error}
\BIBentryALTinterwordspacing
A.~Kandala, K.~Temme, A.~D. C{\'{o}}rcoles, A.~Mezzacapo, J.~M. Chow, and J.~M.
  Gambetta, ``Error mitigation extends the computational reach of a noisy
  quantum processor,'' \emph{Nature}, vol. 567, no. 7749, pp. 491--495, Mar
  2019. [Online]. Available: \url{https://doi.org/10.1038%2Fs41586-019-1040-7}
\BIBentrySTDinterwordspacing

\bibitem{PhysRevA.77.012307}
\BIBentryALTinterwordspacing
E.~Knill, D.~Leibfried, R.~Reichle, J.~Britton, R.~B. Blakestad, J.~D. Jost,
  C.~Langer, R.~Ozeri, S.~Seidelin, and D.~J. Wineland, ``Randomized
  benchmarking of quantum gates,'' \emph{Phys. Rev. A}, vol.~77, p. 012307, Jan
  2008. [Online]. Available:
  \url{https://link.aps.org/doi/10.1103/PhysRevA.77.012307}
\BIBentrySTDinterwordspacing

\bibitem{koczor2021exponential}
\BIBentryALTinterwordspacing
B.~Koczor, ``Exponential error suppression for near-term quantum devices,''
  \emph{Phys. Rev. X}, vol.~11, p. 031057, Sep 2021. [Online]. Available:
  \url{https://link.aps.org/doi/10.1103/PhysRevX.11.031057}
\BIBentrySTDinterwordspacing

\bibitem{larose2022mitiq}
\BIBentryALTinterwordspacing
R.~LaRose, A.~Mari, S.~Kaiser, P.~J. Karalekas, A.~A. Alves, P.~Czarnik,
  M.~El~Mandouh, M.~H. Gordon, Y.~Hindy, A.~Robertson, P.~Thakre, M.~Wahl
  \emph{et~al.}, ``Mitiq: {A} software package for error mitigation on noisy
  quantum computers,'' \emph{{Quantum}}, vol.~6, p. 774, Aug 2022. [Online].
  Available: \url{https://doi.org/10.22331/q-2022-08-11-774}
\BIBentrySTDinterwordspacing

\bibitem{larose2022error}
\BIBentryALTinterwordspacing
R.~LaRose, A.~Mari, V.~Russo, D.~Strano, and W.~J. Zeng, ``Error mitigation
  increases the effective quantum volume of quantum computers,'' \emph{arXiv
  preprint arXiv:2203.05489}, 2022. [Online]. Available:
  \url{https://doi.org/10.48550/arXiv.2203.05489}
\BIBentrySTDinterwordspacing

\bibitem{li2017efficient}
\BIBentryALTinterwordspacing
Y.~Li and S.~C. Benjamin, ``Efficient variational quantum simulator
  incorporating active error minimization,'' \emph{Phys. Rev. X}, vol.~7, p.
  021050, Jun 2017. [Online]. Available:
  \url{https://link.aps.org/doi/10.1103/PhysRevX.7.021050}
\BIBentrySTDinterwordspacing

\bibitem{mari2021extending}
\BIBentryALTinterwordspacing
A.~Mari, N.~Shammah, and W.~J. Zeng, ``Extending quantum probabilistic error
  cancellation by noise scaling,'' \emph{Phys. Rev. A}, vol. 104, p. 052607,
  Nov 2021. [Online]. Available:
  \url{https://link.aps.org/doi/10.1103/PhysRevA.104.052607}
\BIBentrySTDinterwordspacing

\bibitem{mcclean2020decoding}
\BIBentryALTinterwordspacing
J.~R. McClean, Z.~Jiang, N.~C. Rubin, R.~Babbush, and H.~Neven, ``Decoding
  quantum errors with subspace expansions,'' \emph{Nature Communications},
  vol.~11, no.~1, Jan 2020. [Online]. Available:
  \url{https://doi.org/10.1038%2Fs41467-020-14341-w}
\BIBentrySTDinterwordspacing

\bibitem{McClean2017}
\BIBentryALTinterwordspacing
J.~R. McClean, M.~E. Kimchi-Schwartz, J.~Carter, and W.~A. de~Jong, ``Hybrid
  quantum-classical hierarchy for mitigation of decoherence and determination
  of excited states,'' \emph{Phys. Rev. A}, vol.~95, p. 042308, Apr 2017.
  [Online]. Available:
  \url{https://link.aps.org/doi/10.1103/PhysRevA.95.042308}
\BIBentrySTDinterwordspacing

\bibitem{10025519}
\BIBentryALTinterwordspacing
B.~McDonough, A.~Mari, N.~Shammah, N.~T. Stemen, M.~Wahl, W.~J. Zeng, and P.~P.
  Orth, ``Automated quantum error mitigation based on probabilistic error
  reduction,'' in \emph{2022 IEEE/ACM Third International Workshop on Quantum
  Computing Software (QCS)}.\hskip 1em plus 0.5em minus 0.4em\relax Los
  Alamitos, CA, USA: IEEE Computer Society, Nov 2022, pp. 83--93. [Online].
  Available:
  \url{https://doi.ieeecomputersociety.org/10.1109/QCS56647.2022.00015}
\BIBentrySTDinterwordspacing

\bibitem{Ni_2023}
\BIBentryALTinterwordspacing
Z.~Ni, S.~Li, X.~Deng, Y.~Cai, L.~Zhang, W.~Wang, Z.-B. Yang, H.~Yu, F.~Yan,
  S.~Liu \emph{et~al.}, ``Beating the break-even point with a
  discrete-variable-encoded logical qubit,'' \emph{Nature}, vol. 616, no. 7955,
  pp. 56--60, Mar 2023. [Online]. Available:
  \url{https://doi.org/10.1038/s41586-023-05784-4}
\BIBentrySTDinterwordspacing

\bibitem{mike_ike_2020}
M.~A. Nielsen and I.~Chuang, \emph{Quantum computation and quantum
  information}.\hskip 1em plus 0.5em minus 0.4em\relax Cambridge University
  Press, 2010.

\bibitem{piveteau2021error}
\BIBentryALTinterwordspacing
C.~Piveteau, D.~Sutter, S.~Bravyi, J.~M. Gambetta, and K.~Temme, ``Error
  mitigation for universal gates on encoded qubits,'' \emph{Phys. Rev. Lett.},
  vol. 127, p. 200505, Nov 2021. [Online]. Available:
  \url{https://link.aps.org/doi/10.1103/PhysRevLett.127.200505}
\BIBentrySTDinterwordspacing

\bibitem{Postler_2022}
\BIBentryALTinterwordspacing
L.~Postler, S.~Heu$\upbeta$en, I.~Pogorelov, M.~Rispler, T.~Feldker, M.~Meth,
  C.~D. Marciniak, R.~Stricker, M.~Ringbauer, R.~Blatt \emph{et~al.},
  ``Demonstration of fault-tolerant universal quantum gate operations,''
  \emph{Nature}, vol. 605, no. 7911, pp. 675--680, May 2022. [Online].
  Available: \url{https://doi.org/10.1038/s41586-022-04721-1}
\BIBentrySTDinterwordspacing

\bibitem{9773204}
\BIBentryALTinterwordspacing
G.~Ravi, K.~N. Smith, P.~Gokhale, A.~Mari, N.~Earnest, A.~Javadi-Abhari, and
  F.~T. Chong, ``{VAQEM}: A variational approach to quantum error mitigation,''
  in \emph{2022 IEEE International Symposium on High-Performance Computer
  Architecture (HPCA)}.\hskip 1em plus 0.5em minus 0.4em\relax Los Alamitos,
  CA, USA: IEEE Computer Society, Apr 2022, pp. 288--303. [Online]. Available:
  \url{https://doi.ieeecomputersociety.org/10.1109/HPCA53966.2022.00029}
\BIBentrySTDinterwordspacing

\bibitem{inproceedings}
\BIBentryALTinterwordspacing
G.~S. Ravi, J.~M. Baker, A.~Fayyazi, S.~F. Lin, A.~Javadi-Abhari, M.~Pedram,
  and F.~T. Chong, ``Better than worst-case decoding for quantum error
  correction,'' in \emph{Proceedings of the 28th ACM International Conference
  on Architectural Support for Programming Languages and Operating Systems,
  Volume 2}, ser. ASPLOS 2023.\hskip 1em plus 0.5em minus 0.4em\relax New York,
  NY, USA: Association for Computing Machinery, 2023, p. 88–102. [Online].
  Available: \url{https://doi.org/10.1145/3575693.3575733}
\BIBentrySTDinterwordspacing

\bibitem{ravi2022boosting}
\BIBentryALTinterwordspacing
G.~S. Ravi, J.~M. Baker, K.~N. Smith, N.~Earnest, A.~Javadi-Abhari, and
  F.~Chong, ``Boosting quantum fidelity with an ordered diverse ensemble of
  clifford canary circuits,'' 2022. [Online]. Available:
  \url{https://doi.org/10.48550/arXiv.2209.13732}
\BIBentrySTDinterwordspacing

\bibitem{Roffe_2019}
\BIBentryALTinterwordspacing
J.~Roffe, ``Quantum error correction: an introductory guide,''
  \emph{Contemporary Physics}, vol.~60, no.~3, p. 226–245, Jul 2019.
  [Online]. Available: \url{http://dx.doi.org/10.1080/00107514.2019.1667078}
\BIBentrySTDinterwordspacing

\bibitem{russo2022testing}
\BIBentryALTinterwordspacing
V.~Russo, A.~Mari, N.~Shammah, R.~LaRose, and W.~J. Zeng, ``Testing
  platform-independent quantum error mitigation on noisy quantum computers,''
  2022. [Online]. Available: \url{https://doi.org/10.48550/arXiv.2210.07194}
\BIBentrySTDinterwordspacing

\bibitem{schultz2022reducing}
\BIBentryALTinterwordspacing
K.~Schultz, R.~LaRose, A.~Mari, G.~Quiroz, N.~Shammah, B.~D. Clader, and W.~J.
  Zeng, ``Impact of time-correlated noise on zero-noise extrapolation,''
  \emph{Phys. Rev. A}, vol. 106, p. 052406, Nov 2022. [Online]. Available:
  \url{https://link.aps.org/doi/10.1103/PhysRevA.106.052406}
\BIBentrySTDinterwordspacing

\bibitem{Sivak_2023}
\BIBentryALTinterwordspacing
V.~V. Sivak, A.~Eickbusch, B.~Royer, S.~Singh, I.~Tsioutsios, S.~Ganjam,
  A.~Miano, B.~L. Brock, A.~Z. Ding, L.~Frunzio \emph{et~al.}, ``Real-time
  quantum error correction beyond break-even,'' \emph{Nature}, vol. 616, no.
  7955, pp. 50--55, Mar 2023. [Online]. Available:
  \url{https://doi.org/10.1038/s41586-023-05782-6}
\BIBentrySTDinterwordspacing

\bibitem{Suzuki_2022_PRXQ}
\BIBentryALTinterwordspacing
Y.~Suzuki, S.~Endo, K.~Fujii, and Y.~Tokunaga, ``Quantum error mitigation as a
  universal error reduction technique: Applications from the {NISQ} to the
  fault-tolerant quantum computing eras,'' \emph{PRX Quantum}, vol.~3, p.
  010345, Mar 2022. [Online]. Available:
  \url{https://link.aps.org/doi/10.1103/PRXQuantum.3.010345}
\BIBentrySTDinterwordspacing

\bibitem{temme2017error}
\BIBentryALTinterwordspacing
K.~Temme, S.~Bravyi, and J.~M. Gambetta, ``Error mitigation for short-depth
  quantum circuits,'' \emph{Phys. Rev. Lett.}, vol. 119, p. 180509, Nov 2017.
  [Online]. Available:
  \url{https://link.aps.org/doi/10.1103/PhysRevLett.119.180509}
\BIBentrySTDinterwordspacing

\bibitem{RevModPhys.87.307}
\BIBentryALTinterwordspacing
B.~M. Terhal, ``Quantum error correction for quantum memories,'' \emph{Rev.
  Mod. Phys.}, vol.~87, pp. 307--346, Apr 2015. [Online]. Available:
  \url{https://link.aps.org/doi/10.1103/RevModPhys.87.307}
\BIBentrySTDinterwordspacing

\bibitem{PhysRevA.90.062320}
\BIBentryALTinterwordspacing
Y.~Tomita and K.~M. Svore, ``Low-distance surface codes under realistic quantum
  noise,'' \emph{Phys. Rev. A}, vol.~90, p. 062320, Dec 2014. [Online].
  Available: \url{https://link.aps.org/doi/10.1103/PhysRevA.90.062320}
\BIBentrySTDinterwordspacing

\bibitem{PhysRevLett.129.030501}
\BIBentryALTinterwordspacing
Y.~Zhao, Y.~Ye, H.-L. Huang, Y.~Zhang, D.~Wu, H.~Guan, Q.~Zhu, Z.~Wei, T.~He,
  S.~Cao \emph{et~al.}, ``Realization of an error-correcting surface code with
  superconducting qubits,'' \emph{Phys. Rev. Lett.}, vol. 129, p. 030501, Jul
  2022. [Online]. Available:
  \url{https://link.aps.org/doi/10.1103/PhysRevLett.129.030501}
\BIBentrySTDinterwordspacing

\end{thebibliography}

\end{document}